%% file: main_DAC.tex
\RequirePackage{rotating}
\documentclass[sigconf,9pt]{acmart}

\copyrightyear{2026}
\acmYear{2026}
\setcopyright{cc}
\setcctype{by}
\acmConference[DAC '26]{63rd ACM/IEEE Design Automation Conference}{July 26--29, 2026}{Long Beach, CA, USA}
\acmBooktitle{63rd ACM/IEEE Design Automation Conference (DAC '26), July 26--29, 2026, Long Beach, CA, USA}
\acmDOI{10.1145/3770743.3803980}
\acmISBN{979-8-4007-2254-7/2026/07}

\usepackage{tikz}
\usepackage{amsmath}
\usepackage{listings}
\usepackage{xcolor}
\usepackage{rotating}
\definecolor{codegreen}{rgb}{0,0.6,0}
\definecolor{codegray}{rgb}{0.5,0.5,0.5}
\definecolor{codepurple}{rgb}{0.58,0,0.82}
\definecolor{backcolour}{rgb}{0.95,0.95,0.92}

\lstdefinestyle{mystyle}{
	language=c++,
	basicstyle=\ttfamily,
	breaklines=true,
	keywordstyle=\color[RGB]{40,40,255},
	morekeywords={},
	emph={self},
    emphstyle=\bfseries\color{Rhodamine},
    commentstyle=\itshape\color{black!50!white},
    stringstyle=\bfseries\color{PineGreen!90!black},
    columns=flexible,
}

\usepackage{subfigure}
\usepackage{algorithm}  
\usepackage{algorithmicx} 
\usepackage[noend]{algpseudocode}
\usepackage{makecell}
\usepackage{enumitem}
\usepackage{xspace}
\setlist[itemize]{leftmargin=*}
\setlist[enumerate]{leftmargin=*}

\newcommand{\stitle}[1]{\vspace{0.05in}\noindent\textbf{#1}}

\newcommand{\revise}[1]{#1}

\newcommand{\sys}{SmartSwap\xspace}
\begin{document}

\title{\sys: Swap-Based Memory Optimization for LLM Training under Varying Operator Sequences}
\author{\large Zibo Wang$^{1}$, Yuhang Zhou$^{1}$, Zhibin Wang$^{1}$*, Shipeng Li$^{1}$, Xinjing Huang$^{2}$, Chendong Cai$^{2}$, Bingxu Mu$^{2}$, Yuqing Sun$^{2}$, Zhiheng Hu$^{2}$, Bin She$^{2}$, Shu You$^{2}$, Guanghuan Fang$^{2}$, 
Rong Gu$^{1}$, Wanchun Dou$^{1}$, Guihai Chen$^{1}$, Chen Tian$^{1}$ }
\authornote{Zhibin Wang is the corresponding author.}
\affiliation{
\institution{\textit{$^{1}$State Key Laboratory for Novel Software Technology, Nanjing University\qquad$^{2}$Huawei Technologies Co., Ltd}}
\country{}
}

\renewcommand{\shortauthors}{Zibo Wang et al.}
\renewcommand{\shorttitle}{\sys}
\begin{abstract}
    \input{DAC_section/Abstract}
\end{abstract}

\maketitle % should come after the abstract

\input{DAC_section/Sec_1_Introduction}
\input{DAC_section/Sec_2_Background}
\input{DAC_section/Sec_4_Low_overhead_profiling}
\input{DAC_section/Sec_5_Swap_policy_generation}
\input{DAC_section/Sec_6_Executor}
\input{DAC_section/Sec_7_Evaluation}
\input{DAC_section/Sec_9_Conclusion}
\input{DAC_section/Acknowledgments}

\bibliographystyle{plain}
\bibliography{ref-shorten}

\end{document}

%% file: DAC_section/Abstract.tex
The increasing size of large language models (LLMs) has led to a surge in memory requirements during training, often exceeding the capacity of high-bandwidth memory (HBM). Swap-based memory optimization incurs neither accuracy loss nor additional end-to-end overhead when effectively overlapped, thus being an attractive solution. However, existing swap methods assume consistent operator sequences, which is impractical in Eager Mode, where operator sequences can vary across iterations.

We propose \sys, which redesigns the end-to-end process of swap-based memory optimization and is the first work to consider varying operator sequences in Eager Mode. \sys (i) introduces a lightweight online profiler to enable continuous profiling for monitoring operator sequences, (ii) generates effective swap policies with limited operator information, and (iii) optimizes the policy execution module for accurate policy application and better performance. Experimental results demonstrate that \sys reduces profiling overhead by 84.25\%, enables training models up to 4$\times$ larger than hardware memory while adapting to changes in operator sequences, and improves performance by up to 38.94\% compared to recomputation or high-degree parallelism.

\textit{This paper was previously titled “Chameleon: Taming Dynamic Operator Sequences for Memory-Intensive LLM Training.”}

%% file: DAC_section/Sec_1_Introduction.tex
\section{Introduction}\label{sec.intro}

In recent years, large language models (LLMs)~\cite{deepseekai2025deepseekv3technicalreport,deepseekai2025deepseekr1incentivizingreasoningcapability,grattafiori2024llama3herdmodels, kimiteam2025kimik2openagentic,openai2024gpt4technicalreport, bai2025qwen25vltechnicalreport} have emerged as a vibrant research domain due to their remarkable capabilities. To solve more complex problems, researchers have primarily focused on increasing the number of model parameters, which has been proven to be an effective strategy~\cite{kaplan2020scalinglawsneurallanguage, wei2022emergentabilitieslargelanguage}. 

However, the explosive growth in model sizes has made it impossible to complete model training with the limited HBM of a single AI accelerator. There are many methods to alleviate memory limitations, such as parallelization~\cite{li2020pytorchdistributedexperiencesaccelerating, NIPS2014_7d6044e9, NEURIPS2019_093f65e0, 10.1145/3341301.3359646,zhao2023pytorchfsdpexperiencesscaling,295549}, compression~\cite{han2016deepcompressioncompressingdeep, computers12030060}, sparsity~\cite{DBLP:conf/ijcai/LiLTWHLB22, JMLR:v23:21-0998, yuan2025nativesparseattentionhardwarealigned}, recomputation~\cite{chen2016trainingdeepnetssublinear, MLSYS2023_8a27bb69, HONG2025105053}, etc., which are mutually orthogonal and can be used in combination~\cite{10.1145/3178487.3178491, 10.1145/3373376.3378505, DBLP:conf/hpca/RhuOCPKK18, DBLP:conf/ipps/BGK19, DBLP:conf/cluster/ChenHZCHYSC21, DBLP:conf/ics/HuXDLZZMST22, 298555}. %\wzb{better add some citations to support that swap can coorperate with other methods.}
Among them, swap is an ideal memory optimization technique. 
It involves swapping dynamic memory (i.e., activations) to the host DRAM when they are not used for a long period to free up HBM and swapping them back to the device before the next use to ensure uninterrupted training~\cite{DBLP:conf/micro/RhuGCZK16}. 

Mainstream swap techniques~\cite{9355301,273920,10.1145/3458817.3476205,9940581,10.1145/3394486.3406703,DBLP:conf/micro/RhuGCZK16,10.1145/3178487.3178491,10.1145/3373376.3378530,9407112,10.1145/3373376.3378505,DBLP:conf/ics/HuXDLZZMST22,nvidia_nemo} typically operate under the implicit assumption that the operator sequence remains unchanged. This is mainly because these methods were designed for the early popular Graph Mode frameworks, 
such as TensorFlow 1.x~\cite{199317}, MindSpore~\cite{DBLP:series/cir/Lei21}, and others~\cite{chen2015mxnetflexibleefficientmachine, 10.1145/2647868.2654889, 10.1145/2939672.2945397, 10.1145/3620665.3640366}, where the training computation is compiled into a static computation graph that is dispatched once and reused throughout training.
This static structure enables holistic optimizations such as computation fusion and tensor swapping, and many prior works have explored swap-based methods under this paradigm~\cite{10.1145/3178487.3178491, 10.1145/3373376.3378530, 9407112}. However, the complexity of debugging and deployment has led to a gradual shift away from Graph Mode frameworks, motivating the rise of Eager Mode frameworks like Pytorch~\cite{NEURIPS2019_bdbca288} that favor flexibility and ease of development. As of December 2024, only 2\% of open-access implementations of machine learning papers used TensorFlow, while 60\% of implementations adopted PyTorch~\cite{papers_with_code_trends}. 

\begin{figure}[t]
   % \vspace{-0.1in}
  \centering
  \includegraphics[scale=0.38]{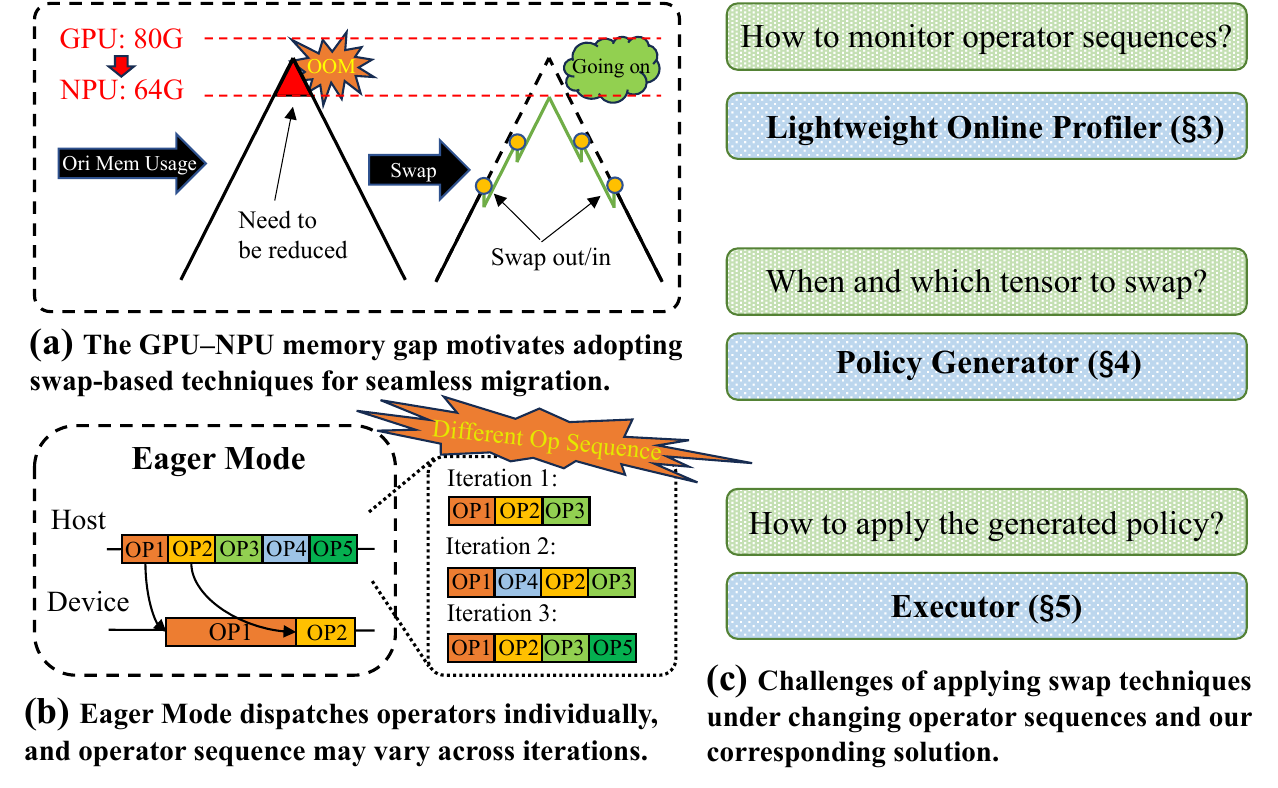}
   \vspace{-0.1in}
  \caption{Leveraging swap-based techniques for seamless migration of PyTorch training from GPU to NPU.}
  \label{fig.e2e}
   \vspace{-0.2in}
\end{figure}

\stitle{Motivating Example: Bridging the GPU–NPU Memory Gap for Seamless PyTorch Training.}
Developers typically prototype and train models in PyTorch on GPUs, leveraging Eager Mode for rapid iteration. Ascend NPUs now provide near drop-in PyTorch support~\cite{pytorch_migration}, but migration often encounters a memory downgrade—for example, A100’s 80 GB HBM versus Ascend 910B’s 64 GB. This 16 GB gap can force changes to model parallelization or device counts, breaking otherwise stable workflows.
We address this gap with a swap-based memory optimization that enables PyTorch workloads to run on lower-capacity NPUs without model refactoring or noticeable performance loss. Although motivated by GPU→NPU migration, the same approach applies to transitions across GPU generations with differing memory capacities.

However, the inherent flexibility of Eager Mode frameworks renders existing swap techniques ineffective.
Prior works~\cite{10.1145/3178487.3178491, 10.1145/3373376.3378530, 9407112, 10.1145/3373376.3378505, DBLP:conf/ics/HuXDLZZMST22} follow a typical \textit{profiling → policy generation → policy application} workflow built on the assumption of consistent operator sequences in Graph Mode frameworks.
They profile operator and tensor behaviors from a single iteration, generate a one-time swap policy that determines which tensors to swap and when, and then directly apply this policy to subsequent iterations.
This design works well under Graph Mode, where operator sequences and tensor identifiers remain stable.
However, in Eager Mode, each operator is dispatched individually to the device and executed sequentially but at different paces with respect to the host. Dynamic features~\cite{pytorch_cond,micikevicius2018mixedprecisiontraining,10.1145/3600006.3613152, 285072, 10.1145/3694715.3695969} bring varying operator sequences across different training iterations, creating conflicts with the prior assumption. This makes the policy generated for earlier sequences invalid and may not only decrease the efficiency of memory optimization but also threaten the reliability of the training system.

To enable efficient swapping in Eager Mode frameworks, we propose Smart-Swap, the first system that systematically supports varying operator sequences in swap-based memory optimization. As in Fig.~\ref{fig.e2e}, \sys also follows the \textit{Profiling → Policy generation → Policy application} workflow, comprising three modules—Lightweight Online Profiler, Policy Generator, and Executor—that address challenges across the full swap workflow, which are detailed as follows.

First, \textit{in the profiling phase, the profiling tools of existing works~\cite{10.1145/3178487.3178491, 10.1145/3373376.3378530, 9407112, 10.1145/3373376.3378505, DBLP:conf/ics/HuXDLZZMST22} cannot meet our requirements for lightweight, online analysis of varying operator sequences.}
Sequence changes can invalidate generated swap policies, causing suboptimal memory use or runtime errors. Existing profilers (i) impose high overhead-e.g., increasing iteration time from 4.9 s to 15.7 s, a 219\% slowdown-and (ii) lack online support, requiring pausing training and defining profiling boundaries manually.
To address this, we implement a \textbf{Lightweight Online Profiler (\S\ref{sec.profiler})} that continuously monitors operator sequences with minimal overhead. It supports two modes: Lightweight and Detailed, and a stage-adjusting module that transitions among WarmUp, GenPolicy, and Stable stages, automatically triggering new swap policy generation when sequences change.

\begin{figure}[t]
   \vspace{-0.1in}
  \centering
  \includegraphics[scale=0.34]{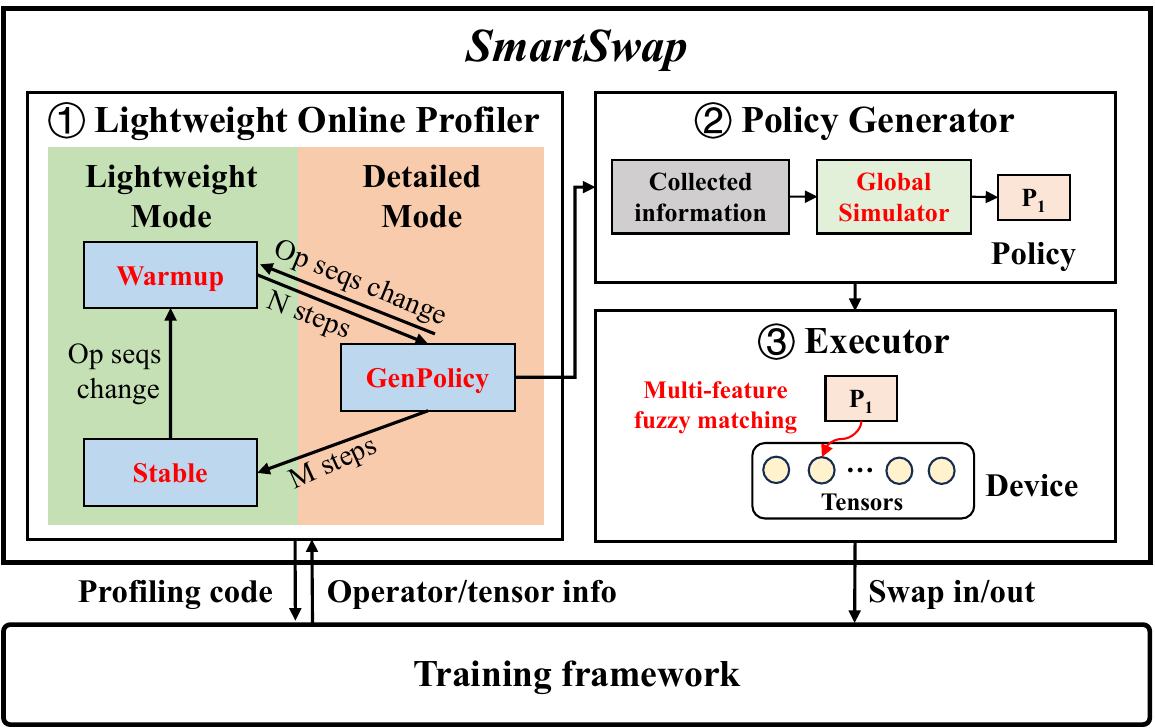}
   \vspace{-0.1in}
  \caption{Overview of \sys and its workflow.}
  \label{fig.e2e}
   \vspace{-0.26in}
\end{figure}

Second, \textit{in the policy generation phase, we face a trade-off between profiling overhead and policy performance.}
To enable continuous online profiling, we exclude high-cost information such as per-operator execution times. However, this missing information complicates the generation of effective swap policies, as accurate timing for pre-triggering swap-in operations typically depends on operator-level timing data. We observe that dividing the operator sequence into evenly sized groups greatly reduces timing variance, allowing the average group time to serve as a low-error proxy for execution time. To balance profiling cost and policy quality, we leverage this insight to design a lightweight \textbf{Policy Generator (\S\ref{sec.policy})} that partitions the operator sequence into logical layers and determines swap operations at this granularity. Moreover, we introduce a global swap simulator to determine precise swap-out and swap-in timings from a global perspective, minimizing performance degradation while maintaining low profiling overhead.

Third, \textit{in the policy application phase, accurately and efficiently applying the generated policy to subsequent iterations is also challenging}, since Eager Mode frameworks recompile and dispatch operators in each iteration without providing unique identifiers to track them across runs.
We design an \textbf{Executor (\S\ref{sec.executor})} that uses multi-feature fuzzy matching to locate corresponding operators and tensors across iterations, while carefully optimizing the overhead of matching.
Moreover, existing frameworks rely on the naive \emph{recordStream} function~\cite{recordStream} for cross-stream memory reuse, but its frequent host–device event synchronization can make the dispatch cost exceed the actual operator execution time, leading to severe host-bound issues. 
To eliminate this bottleneck, we leverage information from the global swap simulator to implement a custom \emph{recordStream} mechanism that replaces host–device synchronization with efficient intra-device stream coordination, thereby avoiding host stalls and further improving overall performance.

We implemented \sys with over 8,700 lines of code, which has been deployed in production for a year and will be open-sourced soon. Experiments on Ascend 910B~\cite{Liao2021Ascend} demonstrate that \sys can adapt to varying operator sequences without causing training errors. In scalability experiments, \sys achieves near-linear scalability with negligible performance degradation. When scaling up the batch size, sequence length, and hidden size of training models, \sys can accommodate models exceeding hardware memory capacity by up to 4$\times$, 4$\times$, and 1.24$\times$ respectively. Moreover, \sys can be leveraged to reduce the degree of parallelism or serve as an alternative to recomputation, achieving up to 38.94\% performance improvement.

%% file: DAC_section/Sec_2_Background.tex
\section{Background}\label{sec.background}

In Graph Mode frameworks, the operator sequence remains fixed, but in Eager Mode frameworks, the integration of large models and dynamic training techniques leads to varying operator sequences, which contradicts the assumptions of existing swap techniques. This variation arises from several factors.

In Eager Mode, each operator is dispatched individually from the host, and dynamic training techniques can alter the sequence. For instance, the computation graph adapts to the model state, executing different operations under varying conditions via conditional branches~\cite{pytorch_cond}, which changes the sequence. Mixed-precision training~\cite{micikevicius2018mixedprecisiontraining}, which adjusts the loss scale for convergence, can shorten the operator sequence if an optimizer update is skipped. Similarly, on-the-fly validation can extend the sequence as it initiates validation at specific stages. Parallel training techniques such as elastic training~\cite{10.1145/3600006.3613152, 285072} or parallelism hot switching~\cite{10.1145/3694715.3695969} also contribute to sequence changes. In practice, operator sequence changes are often observed, especially due to adjustments in loss scale.

When the operator sequence changes, previous swap policies become invalid, leading to issues such as: (i) undersized tensor swaps, failing to prevent Out-of-Memory (OOM) errors; (ii) misaligned lifetimes, where swap timings do not match actual tensor lifetimes, causing suboptimal memory usage; (iii) runtime errors if tensors are not swapped back before their next use, resulting in crashes. This unpredictability not only reduces memory optimization efficiency but also threatens the reliability of the training system.

%% file: DAC_section/Sec_4_Low_overhead_profiling.tex
\section{Lightweight Online Profiler} \label{sec.profiler}
As mentioned in \S\ref{sec.intro}, existing profilers, such as the PyTorch built-in profiler, incur significant overhead, causing substantial performance degradation when used for continuous profiling. They also lack online support, requiring training to pause and hardcode profiling configurations, which is impractical for real-world training. Without continuous online profiling, it is impossible to track and adapt to changes in operator sequences. To address this, we develop a lightweight online profiler that inserts hooks at the operator dispatch point~\cite{opcommand}, operating in either Detailed or Lightweight mode based on whether a new swap policy is needed.

\stitle{Lightweight Mode: } When no new policy is needed, the profiler only collects operators from the current iteration. Inspired by tokenization~\cite{sennrich2016neuralmachinetranslationrare}, we assign an integer to each operator and represent the sequence as an integer tensor. By comparing tensors across iterations, we efficiently detect sequence changes with minimal overhead. The profiler adapts to sequence changes through a stage-adjusting module (Algo.~\ref{alg.stage_adjustment}). The Executor’s multi-feature fuzzy matching in \S\ref{sec.executor} allows \sys to handle minor variations, switching to the WarmUp stage only if the sequence length changes by more than 5\% or the cosine similarity drops below 95\%. Fig.~\ref{fig.e2e} shows the stages: WarmUp does nothing, GenPolicy generates and executes policies, and Stable reuses existing policies.
 \vspace{-0.08in}
\begin{algorithm}[h]
    \caption{Algorithm of Stage Adjusting.}
    \label{alg.stage_adjustment}
    \renewcommand{\algorithmicrequire}{\textbf{Input:}}
    \renewcommand{\algorithmicensure}{\textbf{Output:}}
    \begin{algorithmic}[1]
        \Require \textit{OpSeq}: Operator sequence represented as an integer tensor; \textit{m, n}: Iterations with stable operator sequence before transferring to GenPolicy stage/Stable stage
        \Ensure Stage
        \State static StableStep $\gets$ 0 \Comment{All static variables are}
        \State static PrevOpSeq $\gets$ OpSeq \Comment{initialized only once}
        \State static PrevStage $\gets$ \textit{WarmUp} \Comment{ at the very beginning.}
        \If {diff(len(OpSeq), len(PrevOpSeq)) < 5\% \textbf{and} \\
            \;\;CosineSimilarity(OpSeq, PrevOpSeq) > 95\%}
        \State StableStep $\gets$ StableStep + 1
            \If {PrevStage is \textit{WarmUp}$\And$StableStep > m}
            \State Stage, StableStep $\gets$ \textit{GenPolicy}, 0
            \ElsIf {PrevStage is \textit{GenPolicy}$\And$StableStep > n}
            \State Stage $\gets$ \textit{Stable}
            \EndIf
        \Else
        \State Stage, StableStep $\gets$ \textit{WarmUp}, 0
        \EndIf
        \State PrevStage, PrevOpSeq $\gets$ Stage, OpSeq
    \end{algorithmic}
\end{algorithm}

\vspace{-0.14in}
\stitle{Detailed Mode:} During the GenPolicy stage, \sys switches to Detailed mode, collecting essential data for policy generation with low overhead, including operator names, input and output tensor arrays, and iteration durations. For each tensor, we gather its pointer (data\_ptr), type, usage count, and call stack, which are used for subsequent identification.

In PyTorch, the host dispatches operators asynchronously, with the host and device progressing at different paces. Collecting execution times requires heavy profiling tools like NVIDIA CUPTI~\cite{cupti} or Huawei AscendCL~\cite{acl_profiling}, which generate large amounts of performance data on the device and incur high costs in data transfer and computation. To avoid this overhead, we omit execution time collection and generate swap policies based on operator sequences and iteration durations, as described in \S\ref{sec.policy}.

In addition to operator and tensor data, the profiler tracks memory usage during operator execution. When a swap occurs, the profiler logs swap details, including location, tensor size, and other relevant information. This data allows us to reconstruct memory usage without swaps and use it in policy generation.

%% file: DAC_section/Sec_5_Swap_policy_generation.tex
\section{Policy Generator} \label{sec.policy}
The swap policy determines both memory savings and performance overhead. This section describes how our policy generator uses profiling information to construct an efficient swap policy.

\begin{figure}[t]
   \vspace{-0.14in}
  \centering
  \includegraphics[scale=0.28]{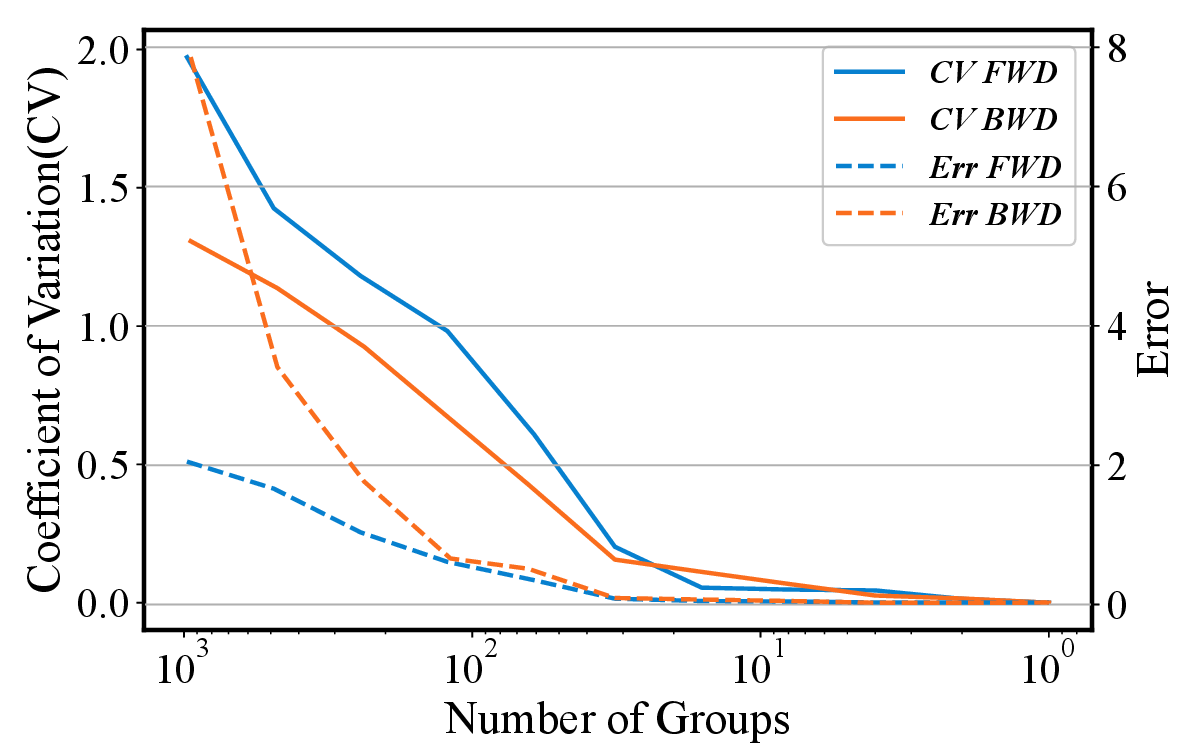}
   \vspace{-0.1in}
  \caption{The relationship between the number of groups and (1) the CV of total execution time per group, and (2) the error of using the time calculated by Eq.(\ref{eq.T_group}) for each group.}
  \label{fig.group_CV_err}
   \vspace{-0.15in}
\end{figure}
\subsection{Observation and Assumption}\label{sec.assumption}
As explained in \S\ref{sec.profiler}, our profiler does not collect per-operator execution times. Instead, we leverage the observation that dividing the operator sequence into evenly sized groups significantly reduces execution-time variance across groups, enabling effective policy generation without requiring exact operator timings.

We trained a 32-layer Llama2 model and profiled operator execution using the PyTorch profiler, analyzing forward and backward propagation separately. For forward propagation, we evenly grouped operators by execution order and measured variation in each group's total execution time using the coefficient of variation (CV). As shown by the blue solid line in Fig.~\ref{fig.group_CV_err}, CV decreases as groups grow, indicating reduced variation in group execution time. Once the group count drops to 32 or fewer, CV approaches zero because each group includes at least one full transformer layer, whose computations are structurally identical. Backward propagation shows a similar pattern. This demonstrates that evenly grouping operators significantly reduces execution-time variation, and—since most modern LLMs are built by stacking similar blocks—this property is expected to generalize. Our policy generator leverages this by grouping operators in forward and backward phases and estimating each group’s execution time with Eq.~(\ref{eq.T_group}).

\vspace{-0.12in}
\begin{align} \label{eq.T_group}
    \overline{T_{group}}=\frac{T_{iter}}{N_{iter}}\times N_{group}.
\vspace{-0.12in}
\end{align}
In this equation, $\overline{T_{group}}$ is the estimated group execution time, $T_{iter}$ is the iteration time, $N_{iter}$ is the number of operators per iteration, and $N_{group}$ is the number of operators per group. The dashed line in Fig.~\ref{fig.group_CV_err} shows the estimation error, which remains low as long as the group count does not exceed the model’s layer count. This confirms the reliability of our grouping-based estimation, and we refer to these groups as \emph{logical layers}.

\subsection{Memory Reduction List (MRL)}\label{sec.memory_reduction_list}
To generate effective swap policies, \sys requires a clear optimization objective: reducing peak memory usage to stay within hardware limits and prevent OOM. Using data collected by the profiler, we reconstruct the memory usage without swaps and build a \emph{memory reduction list (MRL)} by identifying stages where memory exceeds the hardware limit. For each operator in these stages, we create a \emph{memory reduction entry (MRE)} specifying how much memory must be reduced at that point. Since training follows a repeated execution pattern, the allocation–usage–deallocation order remains consistent as long as the operator sequence is unchanged, allowing the MRL to be reused. When the sequence changes and a new policy is needed, we rebuild the MRL using newly collected memory traces, as outlined in Algo.~\ref{alg.policy_generation}.

\subsection{Candidate List (CL)}\label{sec.candidate_list}
During training, although any tensor can be swapped in principle, many swaps provide no memory benefit. For example, tensors used only in the forward phase have lifespans that never overlap with peak memory usage; swapping them wastes PCIe bandwidth and sacrifices opportunities to swap tensors that actually reduce peak memory. Tensors that are too small also lead to poor bandwidth utilization and limited benefit. Thus, we exclude tensors whose lifespans do not overlap with peak memory periods and construct a \emph{candidate list (CL)} from the rest. For each candidate, we compute a score using Eq.~(\ref{eq.score}):
\begin{align} \label{eq.score}
Score = \hat{N_{MRE}} + C\times \hat{S}.
\end{align}
Here, $\hat{N_{MRE}}$ is the normalized number of MREs between the tensor’s last forward use and first backward use, $\hat{S}$ is its normalized size, and $C$ controls their relative importance. Intuitively, larger tensors and those that cover more MRE yield greater potential memory reduction and should be prioritized during policy generation.

\subsection{Simulator}\label{sec.simulator}
We introduce a simulator to accomplish two tasks: determining the time to pre-trigger swap-in and calculating the time at which swap-out is completed. As described in \S\ref{sec.assumption}, we evenly divided the forward operator sequence into logical layers, as well as the backward operator sequence. Within the simulator, each logical layer is represented by a data structure that records key attributes, including the starting operator ID, layer type (forward, backward, or optimizer), assigned swap candidates, and the remaining time available for swap operations within the layer. 

\subsubsection{Pre-trigger Swap-in} \label{sec.simulator.pre-trigger}
Determining the timing of each swap operation is crucial. We first focus on swap-in, which is more complex due to dependencies on the tensor and the inherent host-to-device transfer delay. To avoid execution stalls, swap-ins must be carefully pre-triggered, which we achieve using our simulator.

The simulator processes the CL in descending score order. For each candidate, the required swap-in time is
\begin{align} \label{eq.t_swap}
T_{swap}=S/B,
\end{align}
where $S$ is the tensor size and $B$ is the host-device bandwidth. Starting from the logical layer where the tensor is first used in the backward phase, the simulator searches backward for a layer with $T_{remaining} > T_{swap}$, indicating sufficient time for swap-in without performance degradation. If no suitable layer is found before the peak memory period, the next candidate is considered. If none meet the criteria, we schedule the highest-scoring candidate within its previous logical layer, accepting some latency rather than halting training due to OOM.

Once the swap-in timing is determined, the simulator updates the logical layer by subtracting $T_{swap}$ from $T_{remaining}$, adds the tensor to the layer’s candidate list, and adjusts the MRL by decrementing the tensor’s size from all overlapping memory reduction entries. Swap-in simulation is complete after these updates.

\subsubsection{Swap-out Completion Time} \label{sec.simulator.swap-out}
Since no computation depends on swap-out completion, swaps can be triggered immediately after a tensor’s last forward use. The key is determining when swap-out finishes, which is needed in \S\ref{sec.multi-stream_memory_reuse}. 
The simulator processes each candidate in swap-out order. For each tensor, $T_{swap}$ is computed using Eq.~(\ref{eq.t_swap}). Starting from the logical layer of its last forward use, the simulator searches forward for a layer with $T_{remaining} > T_{swap}$ and records the swap-out completion there. The simulator then updates $T_{remaining}$ and the layer’s candidate list as in \S\ref{sec.simulator.pre-trigger}.

\subsection{Complete Process} \label{sec.policy_gen_comp_process}
The complete policy generation workflow is summarized in Algo.~\ref{alg.policy_generation}. At a high level, our method repeatedly builds an MRL from profiling data, constructs a CL that identifies tensors capable of reducing peak memory, and invokes the simulator to decide viable swap timings. This iterative process continues until all memory reduction requirements are resolved. Finally, swap-out completion times and proactive free events are computed, forming the final swap policy used in subsequent iterations.
\begin{algorithm}[h]
    \caption{Algorithm of Policy Generation.}
    \label{alg.policy_generation}
    \renewcommand{\algorithmicrequire}{\textbf{Input:}}
    \renewcommand{\algorithmicensure}{\textbf{Output:}}
    \begin{algorithmic}[1]
        \Require ProfData
        \Ensure Policy
        \State MRL $\gets$ ConstructMemoryReductionList(ProfData)
        \While {MRL.isNotEmpty()}
        \State CL $\gets$ ConstructCandidateList(ProfData, MRL)
        \If{CL.isNotEmpty()}
        \State PolicyItems $\gets$ Simulator.simulate(CL, MRL)
        \State Policy.extend(PolicyItems)
        \Else
        \State Raise Error
        \EndIf
        \EndWhile
        \State Simulator.SetFreeTime(Policy)
    \end{algorithmic}
\end{algorithm}

%% file: DAC_section/Sec_6_Executor.tex
\vspace{-0.17in}
\section{Executor} \label{sec.executor}
Once the swap policy is generated, the next step is to dispatch the swap operations at the designated positions in the subsequent iteration to realize memory reduction. To ensure precision and efficiency, we developed an \emph{Executor}.
\subsection{Identifying Tensors for Swap}
The policy generator outputs a swap policy specifying which tensors should be swapped in each iteration. Executing this policy requires accurately identifying these tensors at runtime. Although two tensors may be equivalent at the abstract computation-graph level—sharing the same usage patterns and lifecycles—they occupy different physical addresses across iterations and lack persistent identifiers. Thus, Eager Mode frameworks treat them as distinct objects. The same difficulty applies to operators.

While the adaptive stage-switching mechanism in \S\ref{sec.profiler} handles major operator-sequence changes by regenerating policies, we avoid frequent regeneration. The Executor performs \emph{multi-feature fuzzy matching} to locate target tensors, which can adapt to minor sequence variations. It matches tensors using features such as operator name, call stack, and data type, providing robustness to small sequencing shifts. However, naïvely comparing every runtime tensor against every tensor in the swap policy would incur substantial host-side overhead, slowing operator dispatch and causing host-bound performance degradation. To avoid this, we implement several optimizations that allow our fuzzy matching to rely solely on integer comparisons, eliminating expensive operations such as string comparisons.

\begin{figure}[t]
   \vspace{-0.1in}
  \centering
  \includegraphics[scale=0.36]{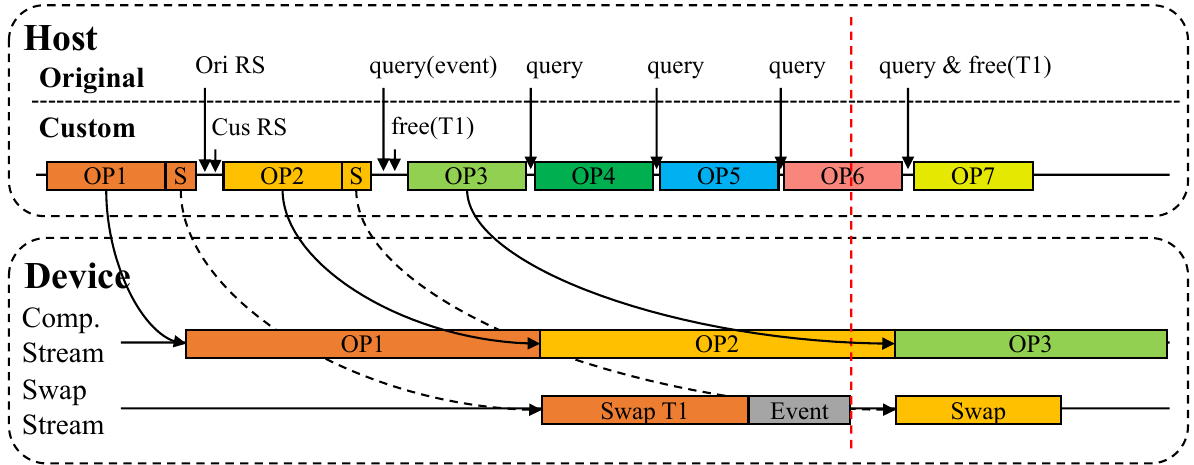}
   \vspace{-0.06in}
  \caption{Comparison of original and custom RecordStream.}
  \label{fig.custom_rs}
   \vspace{-0.28in}
\end{figure}
\subsection{Multi-stream Memory Reuse} \label{sec.multi-stream_memory_reuse}

In Eager Mode frameworks, operators on the same stream execute serially, so placing swap operations directly on the compute stream delays computation. A common workaround is to issue swaps on a dedicated stream, but this creates cross-stream memory reuse hazards. PyTorch manages memory allocation and deallocation on the host side, with each stream having its own memory pool, and memory cannot be reused across streams directly. This design leverages sequential execution within a stream to improve allocation efficiency. PyTorch uses reference counting to automatically release tensor memory when its reference count reaches zero~\cite{NEURIPS2019_bdbca288}, ensuring precise memory freeing. As illustrated in Fig.~\ref{fig.custom_rs}, when T1 is being swapped out in the swap stream, its reference count in the compute stream becomes zero, and its device buffer is freed and may be immediately reallocated to OP2, which can overwrite stale data because the two streams run concurrently.

PyTorch’s \emph{recordStream} avoids this hazard by marking tensors as “in use” by another stream and releasing them only after that stream completes. This ensures correctness but has two drawbacks: it prolongs memory lifetimes—delaying reuse due to asynchronous host/device progress—and it triggers frequent event queries, adding host-side overhead and risking host-bound slowdowns.

To overcome these issues, we implement a custom \emph{recordStream} guided by our Simulator. As described in \S\ref{sec.simulator.swap-out}, the Simulator identifies the compute operator active when a swap-out finishes, giving us precise safe-release points. In the example in Fig.~\ref{fig.custom_rs}, it determines that OP2 is running when T1’s swap-out completes; thus we reclaim T1’s memory immediately after dispatching OP2 and allow reuse as early as OP3 (instead of waiting until OP7 as with the original \emph{recordStream}). This approach eliminates the need for host polling and avoids unnecessary delays in memory reuse. This design enables earlier memory reuse, shortens memory lifetimes, and removes host-bound overhead while maintaining correctness.

%% file: DAC_section/Sec_7_Evaluation.tex
\section{Evaluation}

\textbf{\textit{Implementation.}} We implement \sys with over 8,700 lines of Python and C++ code on top of PyTorch-NPU. Our implementation has been successfully deployed in the production environment for the past year, and we are working towards open-sourcing it in the near future.

\noindent\textbf{\textit{Experimental setup.}} Our experiments are conducted on a server equipped with four ARM-based HiSilicon Kunpeng 920 CPUs, 2 TB RAM, and eight Ascend 910B NPUs, each with 64 GB of HBM. We use Compute Architecture for Neural Networks (CANN) 8.0\footnote{https://www.hiascend.com/en/software/cann} and PyTorch version 2.1.0 for our experiments. \revise{For hyperparameters in Algo.~\ref{alg.stage_adjustment}, we empirically set m to 2 and n to 5. With n = 5, \sys generates five different policies and selects the one with the best runtime performance as the long-term policy.}

\begin{figure*}[t]
  \vspace{-0.2in}
  \centering
    \subfigure[Batch size scaling.]{\includegraphics[scale=0.26]{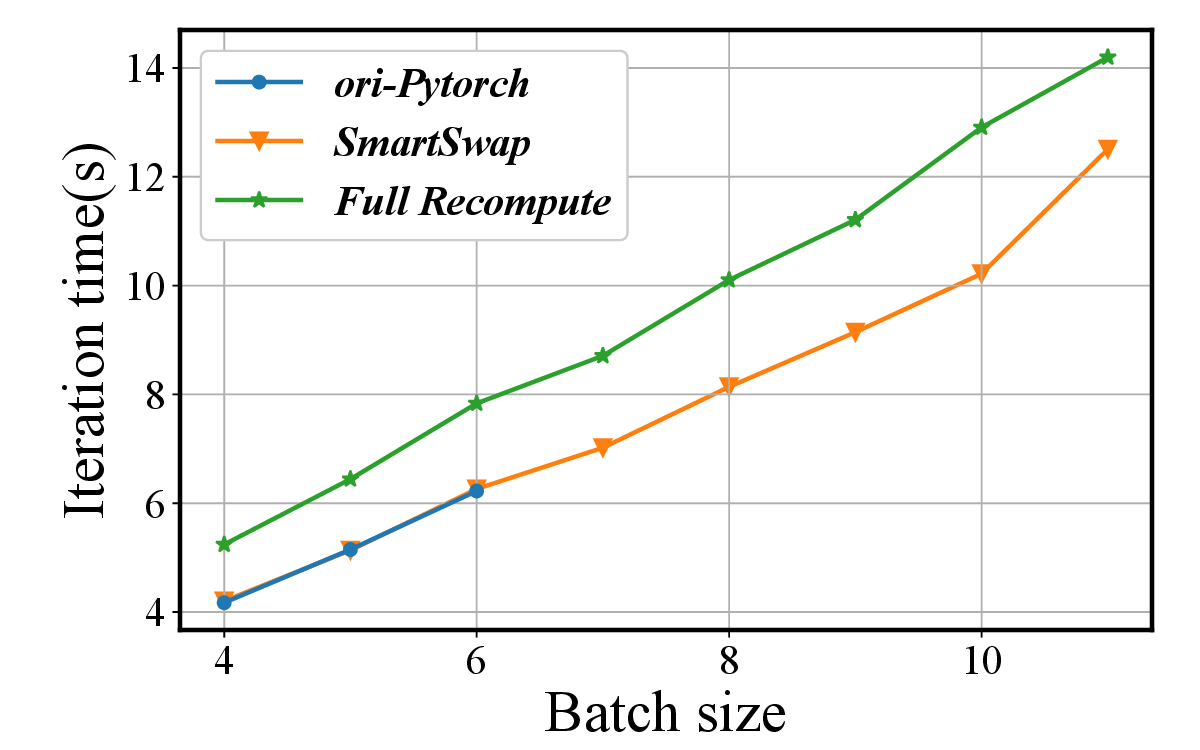}\label{fig.batch_size_extend}}
    \subfigure[Sequence length scaling.]{\includegraphics[scale=0.26]{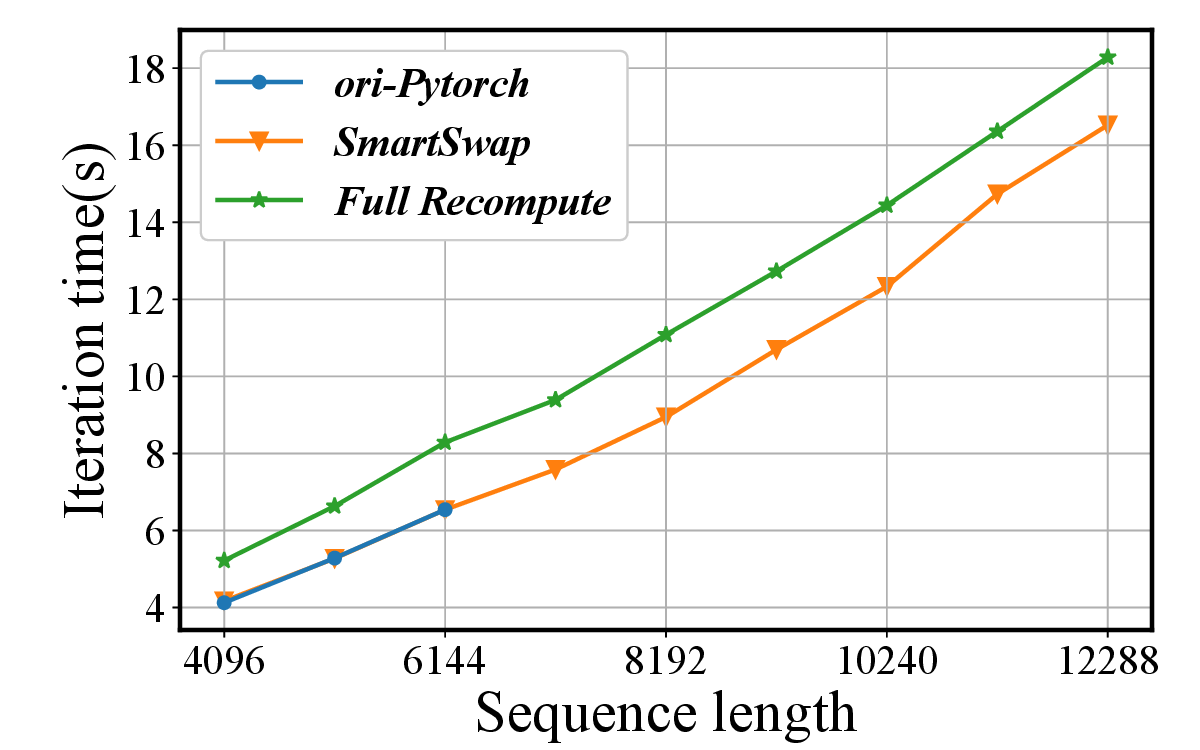}\label{fig.seq_len_extend}}
    \subfigure[Hidden size scaling.]{\includegraphics[scale=0.26]{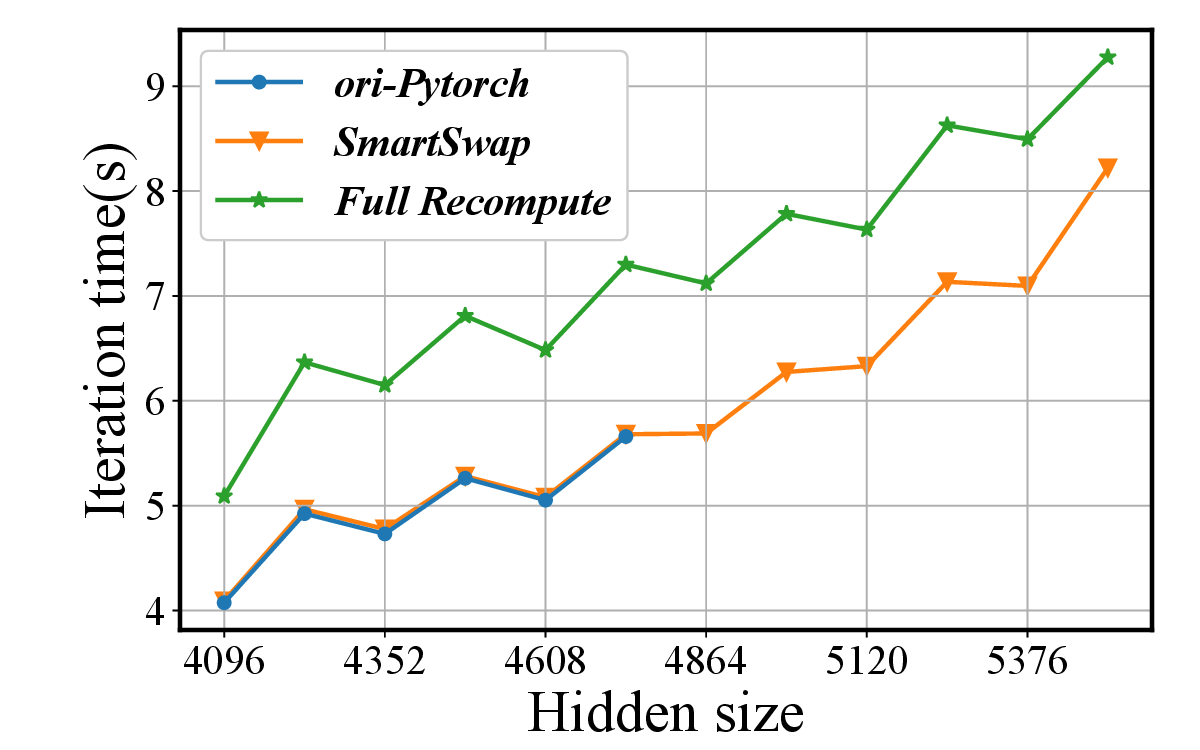}\label{fig.hidden_size_extend}}
  \vspace{-0.14in}
    \caption{Performance under batch size, sequence length, and hidden size scaling.}
    \label{fig.extended_experiment}
\end{figure*}

\subsection{Overall Performance and Scalability}\label{sec.eva.scalability}

To evaluate \sys's performance and scalability, we run experiments along the main model-scaling dimensions: batch size, sequence length, and hidden size, plus a layer-scaling study in \S\ref{sec.eva.custom_rs}. These dimensions represent standard ways of enlarging models in current practice. Using Llama2 as the target model, we assess \sys’s practicality and the maximum model size achievable under each scaling direction.

We record the training performance as the model scales, shown in Fig.~\ref{fig.extended_experiment}. Across all expansion dimensions, \sys consistently meets our targets in \S\ref{sec.intro}. \sys maintains near-linear scaling to 80/64 of PyTorch's maximum and beyond, as swap operations overlap with computation, effectively masking overhead. Compared to full recomputation, \sys improves average performance by 16–19\%, introducing negligible overhead when PyTorch can train without OOM. We further probe \sys’s upper bound. Starting from batch size, layer count, sequence length, and hidden size of 4, 5, 4096, and 4096, we fix three variables and increase the fourth until OOM, recording the last feasible value. Compared to native PyTorch, \sys supports models up to 4$\times$, 1.83$\times$, 4$\times$, and 1.24$\times$ larger along the respective dimensions. Under the same hardware budget, \sys not only trains larger models but also reduces the number of NPUs needed for a given model size, shifting reliance from high-communication TP/PP to DP, improving compute ratio and overall utilization. Across scenarios, \sys provides up to 38.94\% performance improvement.

\subsection{Profiling Overhead}

To evaluate the lightweight profiler, we compare its overhead against the PyTorch profiler on the same Llama2 training task. This task fits on a single NPU without OOM, so \sys produces no policy; thus, all measured time consists solely of computation plus profiling overhead. Each configuration is run five times, and we report the average. Table~\ref{table.prof_overhead} summarizes the results, where \textit{Baseline} is the native PyTorch iteration time.
Our lightweight profiler adds only 0.9\% overhead in low-overhead mode—effectively negligible. Even in detailed mode it incurs just 34.6\% overhead, an 84.25\% reduction relative to the PyTorch profiler’s 219.7\%. These results show that \sys’s online profiler remains lightweight in both modes and is suitable for continuous operator-sequence tracking.

\begin{table}[t]
 \vspace{-0.18in}
    \centering
    \caption{Comparison of profiling overheads.} 
 \vspace{-0.08in}
    \begin{tabular}{|r|r|r|}
        \hline
         & Time (ms) & extra Overhead \\
        \hline
        Baseline & 4,911.1 & / \\
        \hline
        Ours-Low Overhead Mode & 4,952.6 & 0.9\% \\
        \hline
        Ours-Detail Mode & 6,612.0 & 34.6\% \\
        \hline
        Built-in Profiler & 15,699.7 & 219.7\%\\
        \hline
    \end{tabular}
    \label{table.prof_overhead}
 \vspace{-0.15in}
\end{table}

\subsection{Long-term Stability Experiment}\label{sec.long_term}

We evaluate the long-term stability of \sys to ensure that swapping does not compromise training correctness. Using Llama2 scaled to ~80 GB peak memory, we train for 5,000 steps on a single NPU with loss scaling and run online validation every 200 steps. When comparing the loss curves, the curve produced by \sys fully overlaps with that of full recomputation, indicating that \sys preserves training semantics.

We also include a reproduction of Capuchin~\cite{10.1145/3373376.3378505}. Because its code is not available, we implement a PyTorch version following the paper and identify swap targets using an (operator ID, i-th tensor) tuple. Lacking any mechanism to tolerate sequence drift, Capuchin misidentifies tensors and causes the training program to crash in round 201, due to the new operator sequence introduced by on-the-fly validation. In contrast, \sys’s multi-feature fuzzy matching absorbs minor sequence variations and automatically regenerates swap policies when larger changes occur, enabling stable long-running training without runtime errors.

\subsection{Benefit from Custom RecordStream}\label{sec.eva.custom_rs}

\begin{figure}[t]
   \vspace{-0.25in}
  \centering
    \subfigure[Changes in training time as the model size increases.]{\includegraphics[scale=0.23]{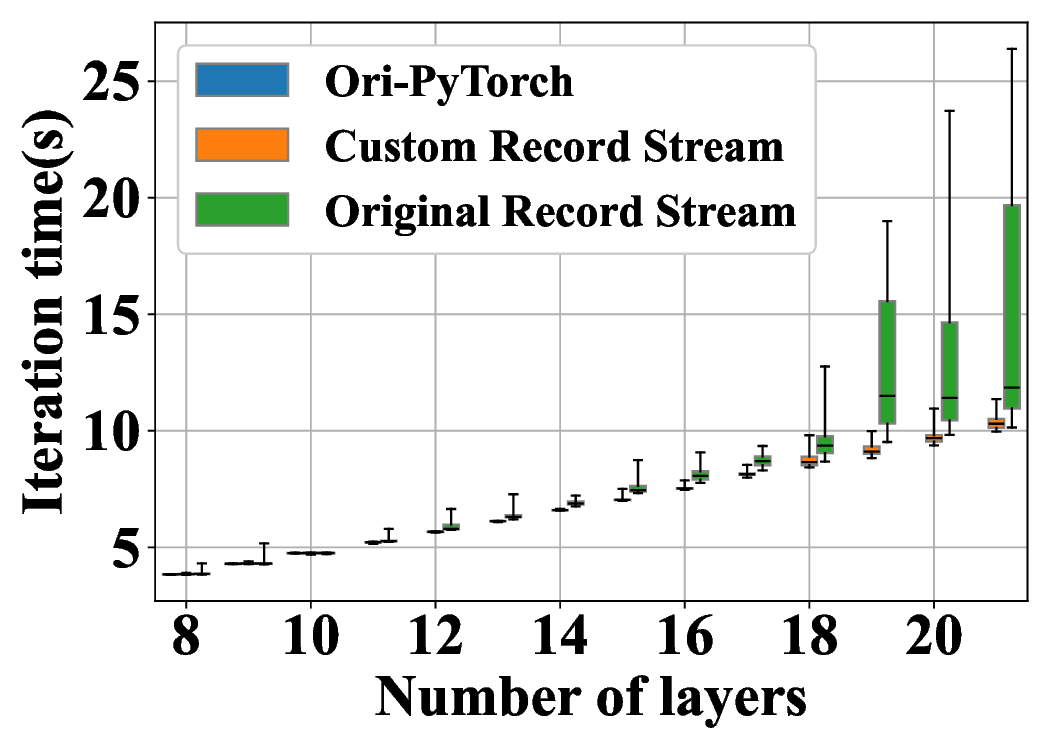}\label{fig.cus_no_cus_comp}}
    \hspace{0.05in}
    \subfigure[Changes in memory block reuse interval as the model size increases.]{\includegraphics[scale=0.23]{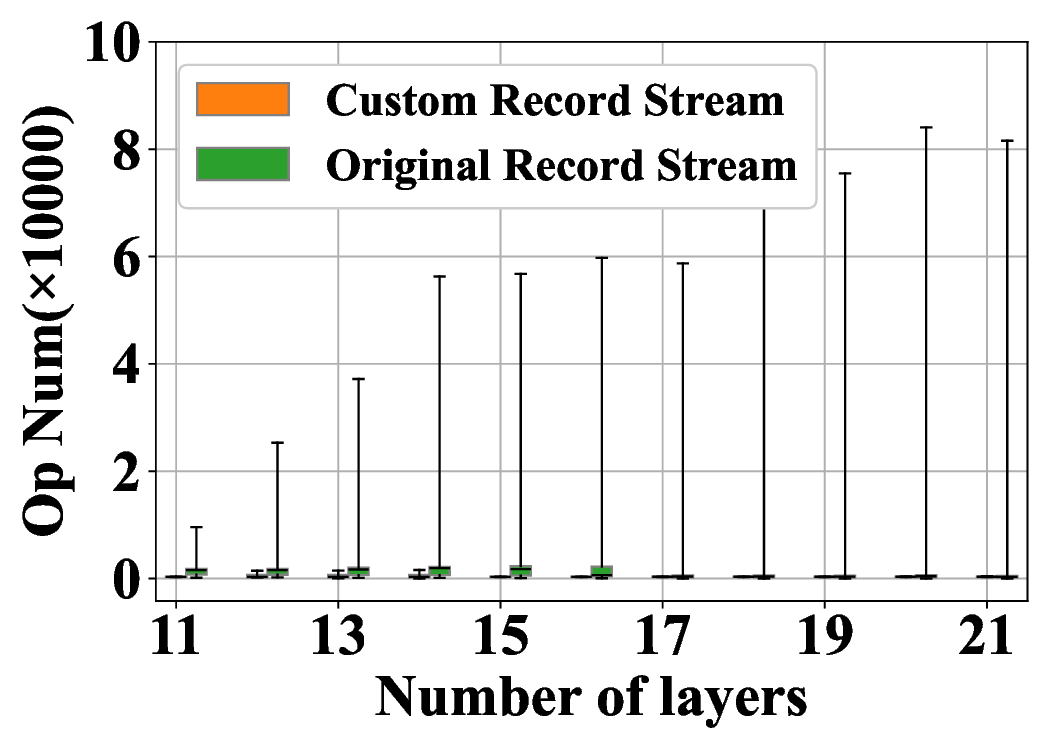}\label{fig.cus_no_cus_rs_op_num}}
   \vspace{-0.1in}
    \caption{Results of comparison experiment between the custom recordStream and the original recordStream.}
   \vspace{-0.2in}
\end{figure}

To validate the benefits of our custom recordStream, we conduct a comparative experiment using Llama2, scaling the model by increasing layer count. Fig.~\ref{fig.cus_no_cus_comp} shows that with the custom recordStream, training time per step scales nearly linearly. In contrast, the original recordStream exhibits significant fluctuations as model size grows. Profiling reveals that frequent event queries increase host-side operator dispatch overhead, causing NPU idle time and host-bound scenario.

To further investigate, we measure the number of operators dispatched between the recordStream call of a memory block and its eventual release back to the memory pool, which we call the memory block reuse interval. As shown in Fig.~\ref{fig.cus_no_cus_rs_op_num}, the original recordStream yields reuse intervals two to three orders of magnitude longer at the tail and 3–4$\times$ larger on average than our custom design. These prolonged intervals force every operator dispatch within them to query event status, substantially increasing CPU overhead and pushing the system into a host-bound scenario, which is the source of the performance fluctuations shown in Fig.~\ref{fig.cus_no_cus_comp}.

%% file: DAC_section/Sec_9_Conclusion.tex
\section{Conclusion}
In this work, we present \sys, a swap-based memory optimization framework redesigned end-to-end to handle the varying operator sequences of Eager Mode frameworks. \sys features a lightweight online profiler for continuous monitoring, generates swap policies with limited operator information, and refines the policy application module for higher accuracy and performance. Experiments show that the profiler reduces overhead by 84.25\%, enabling \sys to track operator sequence changes in real time without introducing training errors. Scalability tests demonstrate near-linear speedup with minimal performance loss, allowing models up to 4$\times$ larger than device memory across multiple scaling dimensions. Moreover, SmartSwap can reduce parallelism requirements or replace recomputation, yielding up to 38.94\% performance improvement. 

%% file: DAC_section/Acknowledgments.tex
\section*{Acknowledgments}
This research is supported by 
the National Natural Science Foundation of China under Grant Numbers 62325205, 62502198 and U25B2035,
and the Postgraduate Research \& Practice Innovation Program of Jiangsu Province (No. KYCX25\_0306).

%% file: ref-shorten.bib
@inproceedings{10.1145/3373376.3378505,
author = {Xuan Peng \emph{et al.}},
title = {Capuchin: Tensor-based GPU Memory Management for Deep Learning},
year = {2020},
isbn = {9781450371025},
url = {https://doi.org/10.1145/3373376.3378505},
doi = {10.1145/3373376.3378505},
booktitle = {ASPLOS 20},
pages = {891–905},
numpages = {15},
location = {Lausanne, Switzerland}
}

@article{papers_with_code_trends,
  author = {Meta AI},
  journal = {\url{https://paperswithcode.com/trends}},
  year = {},
  title = {Accessed: 2024-12. Papers with code trends.}
}

@inproceedings{micikevicius2018mixedprecisiontraining,
  author       = {Paulius Micikevicius \emph{et al.}},
  title        = {Mixed Precision Training},
  booktitle    = {ICLR},
  publisher    = {OpenReview.net},
  year         = {2018},
  url          = {https://openreview.net/forum?id=r1gs9JgRZ},
  bibsource    = {dblp computer science bibliography, https://dblp.org}
}

@inproceedings{10.1145/3600006.3613152,
author = {Insu Jang \emph{et al.}},
title = {Oobleck: Resilient Distributed Training of Large Models Using Pipeline Templates},
year = {2023},
isbn = {9798400702297},
url = {https://doi.org/10.1145/3600006.3613152},
doi = {10.1145/3600006.3613152},
booktitle = {SOSP},
pages = {382–395},
numpages = {14},
location = {Koblenz, Germany}
}

@inproceedings{10.1145/3694715.3695969,
author = {Hao Ge \emph{et al.}},
title = {Enabling Parallelism Hot Switching for Efficient Training of Large Language Models},
year = {2024},
isbn = {9798400712517},
url = {https://doi.org/10.1145/3694715.3695969},
doi = {10.1145/3694715.3695969},
booktitle = {SOSP},
pages = {178–194},
numpages = {17},
location = {Austin, TX, USA}
}

@inproceedings {285072,
author = {John Thorpe \emph{et al.}},
title = {Bamboo: Making Preemptible Instances Resilient for Affordable Training of Large {DNNs}},
booktitle = {NSDI 23},
year = {2023},
isbn = {978-1-939133-33-5},
pages = {497--513},
url = {https://www.usenix.org/conference/nsdi23/presentation/thorpe},
month = apr
}

@article{kaplan2020scalinglawsneurallanguage,
      title={Scaling Laws for Neural Language Models}, 
      author={Jared Kaplan \emph{et al.}},
      year={2020},
      journal={arXiv:2001.08361},
      eprint={2001.08361},
      archivePrefix={arXiv},
      primaryClass={cs.LG},
      url={https://arxiv.org/abs/2001.08361}, 
}

@article{grattafiori2024llama3herdmodels,
      title={The Llama 3 Herd of Models}, 
      author={Aaron Grattafiori \emph{et al.}},
      year={2024},
      eprint={2407.21783},
      journal={arXiv:2407.21783},
      archivePrefix={arXiv},
      primaryClass={cs.AI},
      url={https://arxiv.org/abs/2407.21783}, 
}

@article{li2020pytorchdistributedexperiencesaccelerating,
  author       = {Shen Li \emph{et al.}},
  title        = {PyTorch Distributed: Experiences on Accelerating Data Parallel Training},
  journal      = {Proc. {VLDB} Endow.},
  volume       = {13},
  number       = {12},
  pages        = {3005--3018},
  year         = {2020},
  url          = {http://www.vldb.org/pvldb/vol13/p3005-li.pdf},
  doi          = {10.14778/3415478.3415530},
  bibsource    = {dblp computer science bibliography, https://dblp.org}
}

@inproceedings{NIPS2014_7d6044e9,
 author = {Seunghak Lee \emph{et al.}},
 booktitle = {NeurIPS},
 pages = {},
 title = {On Model Parallelization and Scheduling Strategies for Distributed Machine Learning},
 url = {https://proceedings.neurips.cc/paper_files/paper/2014/file/7d6044e95a16761171b130dcb476a43e-Paper.pdf},
 volume = {27},
 year = {2014}
}

@inproceedings{NEURIPS2019_093f65e0,
 author = {Yanping Huang \emph{et al.}},
 booktitle = {NeurIPS 2019},
 pages = {},
 publisher = {Curran Associates, Inc.},
 title = {GPipe: Efficient Training of Giant Neural Networks using Pipeline Parallelism},
 url = {https://proceedings.neurips.cc/paper_files/paper/2019/file/093f65e080a295f8076b1c5722a46aa2-Paper.pdf},
 volume = {32},
 year = {2019}
}

@inproceedings{10.1145/3341301.3359646,
author = {Deepak Narayanan \emph{et al.}},
title = {PipeDream: generalized pipeline parallelism for DNN training},
year = {2019},
isbn = {9781450368735},
url = {https://doi.org/10.1145/3341301.3359646},
doi = {10.1145/3341301.3359646},
booktitle = {SOSP},
pages = {1–15},
numpages = {15},
location = {Huntsville, Ontario, Canada}
}

@article{zhao2023pytorchfsdpexperiencesscaling,
author = {Yanli Zhao\emph{et al.}},
title = {PyTorch FSDP: Experiences on Scaling Fully Sharded Data Parallel},
year = {2023},
issue_date = {August 2023},
publisher = {VLDB Endowment},
volume = {16},
number = {12},
issn = {2150-8097},
url = {https://doi.org/10.14778/3611540.3611569},
doi = {10.14778/3611540.3611569},
journal = {Proc. VLDB Endow.},
month = aug,
pages = {3848–3860},
numpages = {13}
}

@Article{computers12030060,
AUTHOR = {Zhuo Li\emph{et al.}},
TITLE = {Model Compression for Deep Neural Networks: A Survey},
JOURNAL = {Computers},
VOLUME = {12},
YEAR = {2023},
NUMBER = {3},
ARTICLE-NUMBER = {60},
URL = {https://www.mdpi.com/2073-431X/12/3/60},
ISSN = {2073-431X},
DOI = {10.3390/computers12030060}
}

@inproceedings{DBLP:conf/ijcai/LiLTWHLB22,
  author       = {Yuchao Li \emph{et al.}},
  title        = {Parameter-Efficient Sparsity for Large Language Models Fine-Tuning},
  booktitle    = {IJCAI},
  pages        = {4223--4229},
  year         = {2022},
  url          = {https://doi.org/10.24963/ijcai.2022/586},
  doi          = {10.24963/IJCAI.2022/586},
  bibsource    = {dblp computer science bibliography, https://dblp.org}
}

@article{chen2016trainingdeepnetssublinear,
      title={Training Deep Nets with Sublinear Memory Cost}, 
      author={Tianqi Chen \emph{et al.}},
      year={2016},
      eprint={1604.06174},
      journal={arXiv:1604.06174},
      archivePrefix={arXiv},
      primaryClass={cs.LG},
      url={https://arxiv.org/abs/1604.06174}, 
}

@inproceedings{DBLP:conf/cluster/ChenHZCHYSC21,
  author       = {Ping Chen \emph{et al.}},
  title        = {{CSWAP:} {A} Self-Tuning Compression Framework for Accelerating Tensor
                  Swapping in GPUs},
  booktitle    = {CLUSTER},
  pages        = {271--282},
  publisher    = {{IEEE}},
  year         = {2021},
  url          = {https://doi.org/10.1109/Cluster48925.2021.00019},
  doi          = {10.1109/CLUSTER48925.2021.00019},
  bibsource    = {dblp computer science bibliography, https://dblp.org}
}

@inproceedings{DBLP:conf/ipps/BGK19,
  author       = {Shriram S. B \emph{et al.}},
  title        = {Dynamic Memory Management for GPU-Based Training of Deep Neural Networks},
  booktitle    = {IPDPS},
  pages        = {200--209},
  publisher    = {{IEEE}},
  year         = {2019},
  url          = {https://doi.org/10.1109/IPDPS.2019.00030},
  doi          = {10.1109/IPDPS.2019.00030},
  bibsource    = {dblp computer science bibliography, https://dblp.org}
}

@inproceedings{DBLP:conf/hpca/RhuOCPKK18,
  author       = {Minsoo Rhu \emph{et al.}},
  title        = {Compressing {DMA} Engine: Leveraging Activation Sparsity for Training
                  Deep Neural Networks},
  booktitle    = {HPCA 2018},
  pages        = {78--91},
  year         = {2018},
  url          = {https://doi.org/10.1109/HPCA.2018.00017},
  doi          = {10.1109/HPCA.2018.00017},
  bibsource    = {dblp computer science bibliography, https://dblp.org}
}

@inproceedings{DBLP:conf/ics/HuXDLZZMST22,
  author       = {Zhongzhe Hu \emph{et al.}},
  title        = {MegTaiChi: dynamic tensor-based memory management optimization for
                  {DNN} training},
  booktitle    = {ICS 22},
  pages        = {25:1--25:13},
  publisher    = {{ACM}},
  year         = {2022},
  url          = {https://doi.org/10.1145/3524059.3532394},
  doi          = {10.1145/3524059.3532394},
  bibsource    = {dblp computer science bibliography, https://dblp.org}
}

@inproceedings {298555,
author = {Tailing Yuan \emph{et al.}},
title = {Accelerating the Training of Large Language Models using Efficient Activation Rematerialization and Optimal Hybrid Parallelism},
booktitle = {USENIX ATC},
year = {2024},
isbn = {978-1-939133-41-0},
url = {https://www.usenix.org/conference/atc24/presentation/yuan}
}

@inproceedings{DBLP:conf/micro/RhuGCZK16,
  author       = {Minsoo Rhu \emph{et al.}},
  title        = {vDNN: Virtualized deep neural networks for scalable, memory-efficient
                  neural network design},
  booktitle    = {MICRO},
  pages        = {18:1--18:13},
  year         = {2016},
  url          = {https://doi.org/10.1109/MICRO.2016.7783721},
  doi          = {10.1109/MICRO.2016.7783721},
  bibsource    = {dblp computer science bibliography, https://dblp.org}
}

@inproceedings {199317,
author = {Mart{\'\i}n Abadi \emph{et al.}},
title = {{TensorFlow}: A System for {Large-Scale} Machine Learning},
booktitle = {OSDI 16},
year = {2016},
isbn = {978-1-931971-33-1},
address = {Savannah, GA},
pages = {265--283},
url = {https://www.usenix.org/conference/osdi16/technical-sessions/presentation/abadi},
publisher = {USENIX Association},
month = nov
}

@article{chen2015mxnetflexibleefficientmachine,
      title={MXNet: A Flexible and Efficient Machine Learning Library for Heterogeneous Distributed Systems}, 
      author={Tianqi Chen \emph{et al.}},
      year={2015},
      eprint={1512.01274},
      journal={arXiv:1512.01274},
      archivePrefix={arXiv},
      primaryClass={cs.DC},
      url={https://arxiv.org/abs/1512.01274}, 
}

@inproceedings{10.1145/2647868.2654889,
author = {Yangqing Jia \emph{et al.}},
title = {Caffe: Convolutional Architecture for Fast Feature Embedding},
year = {2014},
isbn = {9781450330633},
url = {https://doi.org/10.1145/2647868.2654889},
doi = {10.1145/2647868.2654889},
booktitle = {MM},
pages = {675–678},
numpages = {4},
location = {Orlando, Florida, USA}
}

@book{DBLP:series/cir/Lei21,
  author       = {Chen Lei},
  title        = {Deep Learning and Practice with MindSpore},
  series       = {Cognitive Intelligence and Robotics},
  publisher    = {Springer},
  year         = {2021},
  url          = {https://doi.org/10.1007/978-981-16-2233-5},
  doi          = {10.1007/978-981-16-2233-5},
  isbn         = {978-981-16-2232-8},
  bibsource    = {dblp computer science bibliography, https://dblp.org}
}

@inproceedings{10.1145/3620665.3640366,
author = {Jason Ansel \emph{et al.}},
title = {PyTorch 2: Faster Machine Learning Through Dynamic Python Bytecode Transformation and Graph Compilation},
year = {2024},
isbn = {9798400703850},
url = {https://doi.org/10.1145/3620665.3640366},
doi = {10.1145/3620665.3640366},
booktitle = {ASPLOS},
numpages = {19},
location = {La Jolla, CA, USA},
}

@inproceedings{10.1145/2939672.2945397,
author = {Frank Seide \emph{et al.}},
title = {CNTK: Microsoft's Open-Source Deep-Learning Toolkit},
year = {2016},
isbn = {9781450342322},
url = {https://doi.org/10.1145/2939672.2945397},
doi = {10.1145/2939672.2945397},
booktitle = {KDD}
}

@inproceedings{10.1145/3178487.3178491,
author = {Linnan Wang \emph{et al.}},
title = {Superneurons: dynamic GPU memory management for training deep neural networks},
year = {2018},
isbn = {9781450349826},
url = {https://doi.org/10.1145/3178487.3178491},
doi = {10.1145/3178487.3178491},
booktitle = {PPoPP 18},
pages = {41–53},
numpages = {13},
location = {Vienna, Austria}
}

@inproceedings{10.1145/3373376.3378530,
author = {Chien-Chin Huang \emph{et al.}},
title = {SwapAdvisor: Pushing Deep Learning Beyond the GPU Memory Limit via Smart Swapping},
year = {2020},
isbn = {9781450371025},
url = {https://doi.org/10.1145/3373376.3378530},
doi = {10.1145/3373376.3378530},
booktitle = {ASPLOS 20},
pages = {1341–1355},
numpages = {15},
location = {Lausanne, Switzerland}
}

@INPROCEEDINGS{9407112,
  author={Jie Ren \emph{et al.}},
  booktitle={HPCA 2021}, 
  title={Sentinel: Efficient Tensor Migration and Allocation on Heterogeneous Memory Systems for Deep Learning}, 
  year={2021},
  volume={},
  number={},
  pages={598-611},
  doi={10.1109/HPCA51647.2021.00057}}

@inproceedings{NEURIPS2019_bdbca288,
 author = {Adam Paszke \emph{et al.}},
 booktitle = {NeurIPS},
 title = {PyTorch: An Imperative Style, High-Performance Deep Learning Library},
 year = {2019}
}

@misc{deepseekai2025deepseekr1incentivizingreasoningcapability,
      title={DeepSeek-R1: Incentivizing Reasoning Capability in LLMs via Reinforcement Learning}, 
      author={DeepSeek-AI \emph{et al.}},
      year={2025},
      eprint={2501.12948},
      journal={arXiv:2501.12948},
      archivePrefix={arXiv},
      primaryClass={cs.CL},
      url={https://arxiv.org/abs/2501.12948}, 
}

@inproceedings {273920,
author = {Jie Ren \emph{et al.}},
title = {{ZeRO-Offload}: Democratizing {Billion-Scale} Model Training},
booktitle = {USENIX ATC 21},
year = {2021},
isbn = {978-1-939133-23-6},
pages = {551--564},
url = {https://www.usenix.org/conference/atc21/presentation/ren-jie},
publisher = {USENIX Association},
month = jul
}

@ARTICLE{9940581,
  author={Jiarui Fang \emph{et al.}},
  journal={IEEE TPDS}, 
  title={Parallel Training of Pre-Trained Models via Chunk-Based Dynamic Memory Management}, 
  year={2023},
  volume={34},
  number={1},
  pages={304-315},
  doi={10.1109/TPDS.2022.3219819}}

@inproceedings {295549,
author = {Ziheng Jiang \emph{et al.}},
title = {{MegaScale}: Scaling Large Language Model Training to More Than 10,000 {GPUs}},
booktitle = {NSDI 24},
year = {2024},
isbn = {978-1-939133-39-7},
address = {Santa Clara, CA},
pages = {745--760},
url = {https://www.usenix.org/conference/nsdi24/presentation/jiang-ziheng},
}

@inproceedings{Liao2021Ascend,
  author={Heng Liao \emph{et al.}},
  booktitle={HPCA}, 
  title={Ascend: a Scalable and Unified Architecture for Ubiquitous Deep Neural Network Computing : Industry Track Paper}, 
  year={2021},
  doi={10.1109/HPCA51647.2021.00071}}

@INPROCEEDINGS{9355301,
  author={Samyam Rajbhandari \emph{et al.}},
  booktitle={SC 20}, 
  title={ZeRO: Memory optimizations Toward Training Trillion Parameter Models}, 
  year={2020},
  volume={},
  number={},
  pages={1-16},
  doi={10.1109/SC41405.2020.00024}}

@inproceedings{10.1145/3458817.3476205,
author = {Samyam Rajbhandari \emph{et al.}},
title = {ZeRO-infinity: breaking the GPU memory wall for extreme scale deep learning},
year = {2021},
isbn = {9781450384421},
url = {https://doi.org/10.1145/3458817.3476205},
doi = {10.1145/3458817.3476205},
booktitle = {SC 21},
articleno = {59},
numpages = {14},
location = {St. Louis, Missouri}
}

@article{nvidia_nemo,
  author = {NVIDIA},
  journal = {\url{https://docs.nvidia.com/nemo-framework/user-guide/latest/nemotoolkit/features/optimizations/cpu_offloading.html}},
  year = {},
  title = {Accessed: 2025-06. NVIDIA NeMo Framework Developer Docs}
}

@inproceedings{10.1145/3394486.3406703,
author = {Jeff Rasley \emph{et al.}},
title = {DeepSpeed: System Optimizations Enable Training Deep Learning Models with Over 100 Billion Parameters},
year = {2020},
isbn = {9781450379984},
url = {https://doi.org/10.1145/3394486.3406703},
doi = {10.1145/3394486.3406703},
booktitle = {KDD},
pages = {3505–3506},
numpages = {2},
location = {Virtual Event, CA, USA}
}

@article{pytorch_cond,
  author = {PyTorch},
  journal = {\url{https://pytorch.org/docs/stable/cond.html}},
  title = {Accessed: 2024-12. Control Flow - Cond.}
}

@article{pytorch_migration,
  author = {Ascend},
  journal = {\url{https://www.hiascend.com/document/detail/zh/Pytorch/700/ptmoddevg/trainingmigrguide/PT_LMTMOG_0014.html}},
  title = {Accessed: 2025-06. Automatic migration.}
}

@article{opcommand,
  author = {Ascend},
  journal = {\url{https://gitee.com/ascend/pytorch/blob/master/torch\_npu/csrc/framework/OpCommand.cpp}},
  title = {Accessed: 2025-06. OpCommand.cpp.}
}

@article{recordStream,
  author = {PyTorch},
  journal = {\url{https://pytorch.org/docs/stable/generated/torch.Tensor.record\_stream.html}},
  title = {Accessed: 2024-12. torch.Tensor.record\_stream.}
}

@article{cupti,
  author = {NVIDIA},
  journal = {\url{https://developer.nvidia.com/cupti}},
  title = {Accessed: 2025-08. NVIDIA CUDA Profiling Tools Interface (CUPTI) - CUDA Toolkit.}
}

@article{acl_profiling,
  author = {Ascend},
  journal = {\url{https://www.hiascend.com/document/detail/zh/canncommercial/82RC1/devaids/Profiling/atlasprofiling_16_0042.html}},
  title = {Accessed: 2025-08. AscendCL Profiling API.}
}

@misc{han2016deepcompressioncompressingdeep,
      title={Deep Compression: Compressing Deep Neural Networks with Pruning, Trained Quantization and Huffman Coding}, 
      author={Song Han \emph{et al.}},
      year={2016},
      journal={arXiv:1510.00149},
      eprint={1510.00149},
      archivePrefix={arXiv},
      primaryClass={cs.CV},
      url={https://arxiv.org/abs/1510.00149}, 
}

@article{JMLR:v23:21-0998,
  author  = {William Fedus \emph{et al.}},
  title   = {Switch Transformers: Scaling to Trillion Parameter Models with Simple and Efficient Sparsity},
  journal = {JMLR},
  year    = {2022},
  volume  = {23},
  number  = {120},
  pages   = {1--39},
  url     = {http://jmlr.org/papers/v23/21-0998.html}
}

@inproceedings{MLSYS2023_8a27bb69,
 author = {Horace He\emph{et al.}},
 booktitle = {MLSys},
 pages = {414--427},
 publisher = {Curan},
 title = {Transcending Runtime-Memory Tradeoffs in Checkpointing by being Fusion Aware},
 url = {https://proceedings.mlsys.org/paper_files/paper/2023/file/8a27bb69950c0b46cdb36d10e5514cc8-Paper-mlsys2023.pdf},
 volume = {5},
 year = {2023}
}

@article{HONG2025105053,
title = {GPU memory usage optimization for backward propagation in deep network training},
journal = {JPDC},
volume = {199},
pages = {105053},
year = {2025},
issn = {0743-7315},
doi = {https://doi.org/10.1016/j.jpdc.2025.105053},
url = {https://www.sciencedirect.com/science/article/pii/S0743731525000206},
author = {Ding-Yong Hong \emph{et al.}}
}

@article{yuan2025nativesparseattentionhardwarealigned,
      title={Native Sparse Attention: Hardware-Aligned and Natively Trainable Sparse Attention}, 
      author={Jingyang Yuan \emph{et al.}},
      year={2025},
      journal={arXiv:2502.11089},
      eprint={2502.11089},
      archivePrefix={arXiv},
      primaryClass={cs.CL},
      url={https://arxiv.org/abs/2502.11089}, 
}

@article{wei2022emergentabilitieslargelanguage,
      title={Emergent Abilities of Large Language Models}, 
      author={Jason Wei \emph{et al.}},
      year={2022},
      journal={arXiv:2206.07682},
      eprint={2206.07682},
      archivePrefix={arXiv},
      primaryClass={cs.CL},
      url={https://arxiv.org/abs/2206.07682}, 
}

@article{sennrich2016neuralmachinetranslationrare,
      title={Neural Machine Translation of Rare Words with Subword Units}, 
      author={Rico Sennrich \emph{et al.}},
      year={2016},
      journal={arXiv:1508.07909},
      eprint={1508.07909},
      archivePrefix={arXiv},
      primaryClass={cs.CL},
      url={https://arxiv.org/abs/1508.07909}, 
}

@article{deepseekai2025deepseekv3technicalreport,
      title={DeepSeek-V3 Technical Report}, 
      author={DeepSeek-AI \emph{et al.}},
      year={2025},
      journal={arXiv:2412.19437},
      eprint={2412.19437},
      archivePrefix={arXiv},
      primaryClass={cs.CL},
      url={https://arxiv.org/abs/2412.19437}, 
}

@article{kimiteam2025kimik2openagentic,
      title={Kimi K2: Open Agentic Intelligence}, 
      author={Kimi Team \emph{et al.}},
      year={2025},
      eprint={2507.20534},
      journal={arXiv:2507.20534},
      archivePrefix={arXiv},
      primaryClass={cs.LG},
      url={https://arxiv.org/abs/2507.20534}, 
}

@article{openai2024gpt4technicalreport,
      title={GPT-4 Technical Report}, 
      author={OpenAI \emph{et al.}},
      year={2024},
      journal={arXiv:2303.08774},
      eprint={2303.08774},
      archivePrefix={arXiv},
      primaryClass={cs.CL},
      url={https://arxiv.org/abs/2303.08774}, 
}

@article{bai2025qwen25vltechnicalreport,
      title={Qwen2.5-VL Technical Report}, 
      author={Shuai Bai \emph{et al.}},
      year={2025},
      journal={arXiv:2502.13923},
      eprint={2502.13923},
      archivePrefix={arXiv},
      primaryClass={cs.CV},
      url={https://arxiv.org/abs/2502.13923}, 
}
